# Exploring Student Interactions with AI-Powered Learning Tools: A Qualitative Study Connecting Interaction Patterns to Educational Learning Theories

Prathamesh Muzumdar (Corresponding author)

College of Business, University of Texas at Arlington

701 S Nedderman Dr, Arlington, TX 76019, USA

E-mail: prathameshmuzumdar85@gmail.com

Sumanth Cheemalapati

College of Business & Information Systems, Dakota State University

820 Washington Ave N, Madison, SD 57042, USA

**Abstract**

With the growing use of artificial intelligence in classrooms and online learning, it has become important to understand how students actually interact with AI tools and how such interactions match with traditional ways of learning. In this study, we focused on how students engage with tools like ChatGPT, Grammarly, and Khan Academy, and tried to connect their usage patterns with well-known learning theories. A small experiment was carried out where undergraduate students completed different learning tasks using these tools, and later shared their thoughts through semi-structured interviews. We looked at four types of interaction: directive, assistive, dialogic, and empathetic—and compared them with learning approaches like behaviorism, cognitivism, constructivism, and humanism. After analyzing the interviews, we found five main themes: Feedback and Reinforcement, Cognitive Scaffolding, Dialogic Engagement, Personalization and Empathy, and Learning Agency. Our findings show that how useful an AI tool feels is not just about its features, but also about how students personally connect with it. By relating these experiences to existing educational theories, we have tried to build a framework that can help design better AI-based learning environments. This work aims to support teachers, EdTech designers, and education researchers by giving practical suggestions grounded in real student experiences.

# 1. Introduction

In recent years, the use of smart technologies in education has grown rapidly, bringing major changes to how students learn and engage with academic content (Jiang et al., 2022). From intelligent tutoring systems to chatbots and large language models, students now regularly interact with software tools that offer help, guidance, and even co-create learning experiences with them. These interactions, which earlier were mostly about giving direct answers or simple feedback, have now become more detailed and conversational. This shift reflects what is commonly called augmented intelligence—a way where technology doesn't replace human thinking but works alongside it to support and strengthen it (Rahman & Watanobe, 2023). As more educational institutions aim to provide personalized learning at scale, these AI-based tools and agents are becoming central to how teaching is planned and delivered in digital learning environments.

Even though more and more educational tools with AI features are being used these days, there's still a major gap in understanding how students' interactions with these software agents match with traditional learning theories (Lo, 2023). Most studies till now have looked at how well these tools work technically or whether students are happy using them. But very few have gone deeper to see if these tools actually support the mental, emotional, and social parts of learning that really matter in the long run. For instance, we still don't know much about how working together with AI tools helps students build knowledge in a constructivist way, or whether such interactions actually encourage students to become more independent in their learning, as suggested by self-determination theory (Kasneci et al., 2023). At the same time, things like trust issues, too much dependence on the tool, moments of confusion, or even feeling mentally overloaded during these interactions haven't been fully explored or explained yet.

In most cases, when AI tools are studied in education, the focus is on numbers—like how long students stay engaged, how much their test scores improve, or how they click through content (Lodge et al., 2023). While these figures are definitely useful, they miss out on the deeper, more personal side of learning—things like how students actually feel, how they reflect, and how they make sense of what the AI tool is doing. These emotional and thoughtful parts of the experience are hard to measure but play a big role in real learning. Learning theories like Vygotsky's socio-cultural theory or Siemens' connectivism remind us that learning happens through context, relationships, and shared meaning-making—not just through content delivery (Noh & Lee, 2020). That's why it's important to look at how students interact with AI tools, and whether these interactions align with or go against the core ideas in these theories.

This study addresses this gap by exploring the following research question:

***How do students experience and interpret their interactions with software agents, and how do these interactions connect to key educational learning theories within their learning journeys?***

To explore this question, our study takes a qualitative route, giving importance to what students themselves share about their experiences. By going through their stories, reflections, and how they interact with AI tools, we try to spot different styles of interaction—like when students delegate tasks to the AI, work together with it, or use it to enhance their learning. These patterns are then linked to concepts from learning theory such as scaffolding, learner control, metacognitive thinking, and feeling socially connected during the learning process. Instead of looking at AI simply as a tool, like earlier research mostly does, this study treats student-AI interactions as more of a two-way relationship that develops within the larger learning system—one that includes both people and technology working together.

Through this work, the paper adds to the growing discussion around the role of intelligent systems in education—focusing not just on what these tools can do, but more importantly, on how students actually experience and make sense of them. It highlights how learners adopt and give meaning to these tools in their own ways. By bringing in student perspectives, the study builds a strong base for educators, curriculum designers, and policymakers who want to bring AI into classrooms in a thoughtful and practical way.

The rest of this paper is organized as follows. Section 2 goes over past research related to intelligent technologies in education, different ways students interact with software agents, and the main learning theories we refer to. Section 3 explains the theoretical framework that helps us look at student experiences more meaningfully. Section 4 shares the qualitative approach we followed, including how we collected and analyzed the data. Section 5 highlights the main findings, grouped under different themes. In Section 6, we connect these findings with educational theories and think about how the role of AI in learning is changing. Finally, Section 7 talks about what this means for actual teaching and design work, and suggests some areas where future research can build on this study.

## 2. Literature Review

The use of intelligent systems in education has brought a big shift in the way students engage with content, teachers, and digital platforms (Surameery & Shakor, 2023). With more tools like AI-powered feedback systems, smart tutoring platforms, and chatbots coming into use, students are now regularly interacting with software agents as part of their daily learning process (Nückles et al., 2020; Ali et al., 2023). These systems are no longer just static tools that push content—they have become active and responsive parts of the learning environment. Because of this, how students think, learn, and take control of their learning is changing in many ways. To properly understand these changes, it becomes important to focus on two key areas: Human-Agent Interaction (HAI) models and the foundational learning theories that explain how people learn.

Behaviorism mainly looks at what can be seen—how people behave—and how learning happens through rewards and repetition. In this view, AI tools that give immediate feedback, rewards, or even game-like features such as scores and badges, act as motivators to help students repeat and master certain tasks (Ali et al., 2023). A good example of this is when tutoring systems change the level or type of question depending on how the student is doing. These designs clearly reflect behaviorist thinking. But at the same time, some experts feel that this approach doesn't really pay attention to what's going on inside the learner's mind—like their thoughts or choices—and may limit their independence in learning (Sullivan et al., 2023).

Cognitivism puts focus on how the mind works—how we process information, solve problems, and build understanding (Alexander & Murphy, 2024). AI tools that help students see difficult ideas more clearly, give timely hints, or let them explore situations through simulations fit well with this approach (Lombardi et al., 2024). Such systems are useful for building mental models, encouraging self-awareness in thinking, and helping learners apply knowledge in new situations. Augmented intelligence tools that guide students through step-by-step problem solving or show them how to think through a complex task are a good example of how this theory comes into practice (Mayer, 2024).

Constructivism, as explained by scholars like Piaget and Vygotsky, sees learning as something that happens actively and in context—where students build their own understanding based on their experiences (Sweller, 2023). AI tools that let learners explore freely, simulate real-life situations, or give timely support act like learning partners or guides in this process (Jalil et al., 2023). When these tools operate within what Vygotsky called the Zone of Proximal Development (ZPD), they can help students by giving just enough help to move forward, and slowly hand over the control so that the learner becomes more independent and confident in their thinking (Bernacki, 2025).

Humanism focuses on the overall growth of the learner—helping them reach their full potential and supporting education that is centered around the individual (Mayer, 2024). From this perspective, AI systems shouldn't just be used to push information; they should also help students become more independent, emotionally balanced, and confident in their personal learning journeys (Sweller, 2023). Tools that let students personalize their learning, give space for reflection, or respond to emotional cues are better suited to humanistic learning. However, many present-day AI tools still struggle to show empathy or support a student's sense of identity, which is a serious gap in their design (Alexander & Murphy, 2024).

Connectivism, which is considered a learning theory suited for the digital era, says that learning happens through networks made up of both people and digital tools (Alexander & Murphy, 2024). In this view, AI systems and software agents become active members of the learner's network, not just background tools. They help learners spot patterns, suggest useful links, and make it easier to access information spread across different platforms (Bernacki, 2025). Features like content recommendations or group collaboration tools show how knowledge today isn't just kept in a person's head—it's spread out across digital systems where it can be accessed and built upon through shared interactions (Mayer, 2024).

While these learning theories provide valuable ways to understand how students learn, there is still a big research gap when it comes to linking specific Human-AI Interaction (HAI) patterns to educational theories (Bernacki, 2025). Most of the current work in this area is more focused on numbers—like whether students get better marks or finish tasks faster—rather than how they actually *feel* and *think* during these interactions (Sweller, 2023). There is very little attention given to how students make sense of their experience, how much control they feel they have, or what emotional responses they go through while working with AI tools (Limna et al., 2023). Also, many studies look at AI tools as if they're just technical support, without seeing them as active teaching partners who can actually shape how learning takes place. This is a major blind spot we hope to address through this study.

This study addresses these gaps by asking the central research question:

How do students experience and interpret their interactions with software agents, and how do these interactions connect to key educational learning theories within their learning journeys?

Through this qualitative study, we are trying to understand how students deal with different kinds of Human-AI Interaction (HAI) patterns and how these experiences connect to their overall learning and development (Miles & Huberman, 1994). Our goal is to link real student experiences to established educational theories and offer a clear pathway between how AI systems are designed and how learning actually happens. The review of literature so far has helped shape a strong foundation for our theoretical framework, which we will explain in the next section. This framework brings together five major learning theories and matches them with key HAI paradigms, allowing us to look at student-AI interactions from multiple educational angles. In doing so, we hope to build a deeper understanding of how AI tools and human learning work hand-in-hand in today's classroom settings.

## 3. Theoretical Framework

With augmented intelligence and automation becoming deeply embedded in today's classrooms, it has become important to understand how students engage with such intelligent tools through proper theoretical lenses (Gomes & Mendes, 2014). In this research, we have built a theoretical framework that brings together Human-Agent Interaction (HAI) models with five key learning theories—Behaviorism, Cognitivism, Constructivism, Humanism, and Connectivism—to help us analyse how learning theories show up in students' real-world interactions with AI tools (Mayer, 2024). HAI paradigms act like a map to show how students relate to and depend on AI systems (Sweller, 2023). These patterns can range from Automation, where the system works fully on its own; to Delegation, where students hand over certain tasks to the system; to Augmentation, where the system supports the student's thinking process; and finally, to Co-Creation, where both the student and AI tool jointly build new understanding or outcomes (Sweller, 2023; Mhlanga, 2023). Each of these modes carries a different level of control for the student, and changes how collaboration between human and machine happens during learning.

Instead of pre-deciding which HAI paradigm fits into which learning theory, this study takes a qualitative route to understand how students themselves view and experience their learning with the help of software agents (Keyvan & Huang, 2022). By conducting interviews, we collected detailed personal accounts that reveal how learners think about the role of AI—whether it's guiding them through a topic, giving timely support, helping them reflect, or simply keeping them engaged. These stories form the heart of our analysis. Rather than forcing a top-down structure, we used these real experiences to inductively figure out how different interactions with AI systems naturally connect—either directly or in subtle ways—to the key ideas behind major learning theories. This bottom-up approach respects students' voices and helps ground the theoretical links in lived realities.

For example, when students talk about doing the same kind of tasks repeatedly and receiving feedback from AI, it often hints at a Behaviorist approach, where learning is seen as a result of reinforcement and observable responses. This is quite common when AI tools give scores, suggest corrections, or provide nudges that push students to keep improving. On the other hand, when learners mention that the AI helped them break down a tough topic or visualize something abstract, they're likely describing something closer to the Cognitivist line of thinking, which gives importance to how we process and organize information mentally (Yilmaz & Yilmaz, 2023). In many cases, students also share experiences where they explored topics on their own, played around with simulations, or solved problems with some AI help—this clearly points towards Constructivist learning, where knowledge gets built through hands-on experiences. If someone brings up how an AI tool seemed to "understand" them, offered personalized tips, or made them feel more confident or free to learn at their own pace, that reflects Humanist ideas, which focus on the learner's personal growth and emotional well-being. Finally, when students talk about sharing work with peers, using AI to discover new resources, or learning through digital communities, it shows a Connectivism perspective, where learning is spread across different platforms and networks—both human and AI-based.

This framework has been kept open and exploratory in nature, rather than being rigid or rule-based. Instead of setting fixed links between learning theories and AI interaction models from the beginning, it allows these connections to emerge naturally from what students share in their own words. The main intention is not to judge AI tools only by their design or functions, but to understand how students actually experience them in their day-to-day learning. How do they see these tools? How do they emotionally connect, use them, or even struggle with them? That is what this study tries to capture. At the same time, this framework is not just for background theory—it has also shaped the way interviews were designed and how we analyzed the responses. It helped us look deeper into how students think, feel, and behave while using AI in their learning journey. Ultimately, it has allowed us to think more seriously about the teaching value of AI—not only in terms of final marks or outcomes, but also in terms of the actual process through which learning happens.

At its core, this theoretical framework aims to address an important gap in the current research—how the design of AI tools connects with foundational learning theories, especially when viewed through the lens of students' real, lived experiences. It treats software agents not just as neutral or passive tools delivering information, but as active contributors in the learning process. These systems play a role in shaping how students come to know, understand, and reflect on what they learn. In today's classrooms, where technology is deeply woven into the fabric of education, such agents are becoming key partners in constructing and supporting knowledge, alongside traditional methods of teaching.

## 4. Methodology

This research uses a qualitative approach to deeply understand how students engage with software agents and how such interactions relate to key educational learning theories (Miles & Huberman, 1994). The main aim is to capture students' personal experiences and perspectives, making qualitative methods most appropriate. While quantitative or mixed-methods designs are often used to track performance or engagement levels, they don't fully capture the nuanced, context-specific ways students make sense of AI tools in learning environments. Hence, this study emphasizes rich, detailed insights over broad statistical generalizations.

Among the various qualitative strategies explored, semi-structured interviews were chosen as the primary method for data collection (Miles & Huberman, 1994). This approach provided the flexibility needed to delve into students' thoughts and emotions, while also offering enough structure to guide discussions around specific interaction styles and theoretical frameworks (Wang et al., 2023). Other techniques, such as think-aloud protocols or diary-based studies, were considered but found to be less practical due to the study's time and resource constraints. Semi-structured interviews offered a balanced approach, combining purposeful questioning with space for students to share personal experiences with AI tools in their learning journey.

For analyzing the data, two qualitative approaches were initially considered: Phenomenological Analysis and Thematic Analysis. Phenomenological Analysis is particularly useful for capturing the depth of a single lived experience by setting aside prior assumptions to uncover the core meaning of that experience. However, this method is less suitable for research that involves comparing experiences across multiple participants or aligning responses with structured theoretical categories. Since this study aims to classify different types of student-AI interactions and connect them to established learning theories—like behaviorism, cognitivism, constructivism, humanism, and connectivism—the underlying philosophy of Phenomenology did not align well with our research goals.

Thematic Analysis, especially following the framework proposed by Braun and Clarke, was selected as the primary method for data interpretation. This approach offers a balanced mix of structure and flexibility, allowing researchers to systematically identify, code, and make sense of recurring patterns across qualitative data (Agbo et al., 2019; Tlili et al., 2023). One of the key strengths of Thematic Analysis is its ability to accommodate both inductive coding—where themes emerge from the data itself—and deductive coding guided by pre-existing theories. This makes it particularly suitable for our study, which seeks to explore diverse student experiences with AI tools while also grounding those narratives in established learning theories. Unlike Phenomenology, Thematic Analysis does not require researchers to completely bracket their theoretical knowledge, which aligns well with our goal of creating a conceptual bridge between Human-Agent Interaction paradigms and pedagogical frameworks.

To sum up, this research employs semi-structured interviews as the core method for gathering insights and uses Thematic Analysis to interpret those insights. This methodological choice is well-suited to our aim of exploring how students perceive, engage with, and learn through interactions with AI-based systems. It also offers a structured yet adaptable framework for identifying meaningful patterns in student narratives, which can then be connected to established pedagogical theories. Through this approach, the study creates a clear and grounded pathway to build theoretical linkages between specific Human-Agent Interaction types and core educational models.

*4.1 Data Collection Instrument*

The semi-structured interview guide used in this study was self-developed and grounded in the theoretical framework connecting Human-AI Interaction paradigms with five established learning theories: Behaviorism, Cognitivism, Constructivism, Humanism, and Connectivism. The questions were designed to elicit student reflections on cognitive, emotional, and behavioral engagement with AI tools. To ensure content validity, the instrument was reviewed by two experts in educational psychology and instructional design. Additionally, a pilot test was conducted with three undergraduate students to confirm clarity and relevance. Feedback from the pilot resulted in minor modifications to question phrasing and sequencing, improving overall reliability and consistency in interpretation.

## 5. Experiment

The experiment for this study was thoughtfully planned to mirror real-life learning situations, where undergraduate students were asked to engage with commonly used AI-powered educational tools while completing specific tasks. Conducted in a controlled lab-like environment, the purpose was to closely observe the nature of their interactions with various software agents and understand how these interactions reflected different teaching and learning styles. The tools selected for the experiment—ChatGPT, Grammarly, and either Khan Academy or Socratic AI—were chosen carefully, as they offer a good mix of interaction modes, from one-way instruction to conversational engagement. These tools are already popular in student learning routines and are known for providing feedback, assisting with content creation, and promoting self-directed learning. The key aim of this experiment was to examine whether and how students' usage of these tools could be matched to well-known learning theories such as behaviorism, cognitivism, constructivism, humanism, and connectivism.

Participants were asked to complete three academic activities; each thoughtfully aligned with a particular learning approach. For the behaviorist condition, students worked on math problems using platforms like Khan Academy or Socratic AI, allowing us to observe how repetition and feedback shaped their responses. The second task, designed from a cognitivist angle, required students to revise a short essay with the help of tools like Grammarly or ChatGPT, focusing on how they processed structured suggestions and corrections. Lastly, in the constructivist-humanist condition, students participated in an open-ended conversation with ChatGPT on topics such as personal learning goals or ethical questions. This was meant to encourage deeper thinking, reflection, and shared meaning-making. All participants completed these tasks one after another, with short breaks and clear instructions between each phase. This structure ensured everyone had a uniform experience while still leaving room for personal interpretation, which was essential for the follow-up interviews that aimed to capture their inner thought processes and learning insights.

## 6. Data Collection

The data collection process was carried out through a structured, task-oriented setup designed to closely examine how students engage with AI-driven educational tools and how such interactions reflect established learning theories. The sessions took place in a university research lab, where all participants used the same set of pre-configured laptops. This arrangement helped maintain uniformity in the technical environment and ensured that differences in interaction were due to student behavior and not device-related issues.

A total of fifteen undergraduate students, belonging to different academic streams, were selected from an introductory course focusing on education and technology. Before starting the study, each student gave informed consent and was given a clear explanation of the experiment process. The activity involved three guided tasks, with each task taking around 10 to 15 minutes. These tasks were designed to reflect typical learning situations where students generally use AI-based tools. The tools chosen for the experiment included ChatGPT, Grammarly, and either Khan Academy or Socratic AI.

The tasks were intentionally mapped to reflect different educational learning theories, as shown in Table 1 below:

Table 1. AI Tools, Tasks, and Associated Learning Theories Table

| Task Description | AI Tool Used | Learning Theory | Interaction Style |
| --- | --- | --- | --- |
| Solve a set of 3–4 math problems | Khan Academy / Socratic AI | Behaviorism | Directive, feedback-based |
| Revise a short essay using AI grammar and content suggestions | Grammarly / ChatGPT | Cognitivism | Assistive, scaffolded |
| Engage in a reflective dialogue on academic challenges | ChatGPT | Constructivism & Humanism | Dialogic, student-centered |

Once the tasks were completed, each student was invited for a semi-structured interview lasting around 30 to 45 minutes. The conversations were audio-recorded with prior permission and later typed out for detailed qualitative study. The purpose of these interviews was to understand how students felt while using each AI tool—what they thought, how they responded emotionally, and whether they felt in control of their learning. Through open-ended questions, students were encouraged to compare their experience with AI to that of learning with a human teacher or guide, and to share whether they felt supported, challenged, or independent during the activity.

During the experiment, field notes were taken regularly to note down students' body language, expressions, and visible reactions—like moments of hesitation, signs of comfort or struggle, or curious engagement. These observations helped add more depth to the interview findings and gave a fuller picture of how students were responding to the AI tools. The notes worked as a supplementary source to strengthen our interpretation of the themes. A quick summary of the key focus areas during the interviews is provided in Table 2.

Table 2. Interview Themes and Types of Insights Collected Table

| Interview Focus Area | Type of Insight Collected |
| --- | --- |
| Perceptions of AI support and usefulness | Cognitive scaffolding, feedback quality, clarity of explanations |
| Emotional responses during interaction | Frustration, motivation, satisfaction, confusion, curiosity |
| Sense of control and learning agency | Perceived autonomy vs. dependency on AI suggestions |
| Comparison with human instruction | Relational and pedagogical differences |
| Interaction interpretation | AI as tutor, peer, coach, or assistant (mapped to HAI paradigms) |

Once all the data were collected, they were anonymized to maintain students' privacy and then sorted for analysis using Thematic Analysis. This method helped us spot common patterns and repeated ideas across what students shared. It also allowed us to link the way students interacted with AI tools to the main educational theories discussed earlier. By combining what we saw during the tasks with what students said in the interviews, we got a well-rounded picture of how learners were understanding and reacting to their experiences with AI-based tools.

## 7. Data Analysis

For analyzing the data, we went ahead with thematic analysis as our main method. This approach gave us the flexibility to dig deep and identify patterns, meanings, and links across what students shared in their interviews. It turned out to be a good fit because we were trying to understand how students made sense of their interactions with AI tools—not just in terms of what they did, but also how they felt and learned during the process. Since the study was exploratory, and we wanted to see how different types of Human-AI interactions connect to educational theories, thematic analysis allowed us to look at the data with an open mind instead of forcing it into fixed boxes or ready-made frameworks.

The audio recordings from the semi-structured interviews were transcribed word-for-word, and all personal identifiers were removed to maintain the confidentiality of participants. Thematic analysis was carried out in six stages: getting familiar with the transcripts, identifying initial codes, looking for common themes, refining these themes, labelling them properly, and finally preparing the detailed analysis report. We used a mix of approaches—some codes naturally came out of what the students said (inductive), while others were based on earlier ideas, we already had about how different types of Human-AI interactions might connect to learning theories (deductive). For instance, when students talked about the AI giving straightforward answers and reinforcing their responses, those bits were grouped under behaviorist themes. On the other hand, when the conversations felt more reflective or exploratory, especially when the AI supported personal thought or goal-setting, those were linked to constructivist or humanistic ideas.

Along with the interview transcripts, field notes taken during the experiment were also examined closely. These notes helped us make sense of non-verbal behaviours like how attentive, confused, or satisfied the students seemed during each task. This kind of observation added depth to our understanding and helped us cross-check the themes we identified from the interviews. For instance, when some students took extra time or seemed unsure before responding to what the AI suggested, they later mentioned in their interviews that they weren't fully convinced the tool matched their learning style. This kind of pattern helped us build a sub-theme around learner agency and how much students trusted AI-driven feedback.

Several overarching themes emerged from the data, including:

- Feedback and Reinforcement,
- Scaffolded Cognitive Support
- Dialogic Engagement
- Personalization and Empathy
- Perceived Learning Agency

Once the themes were finalized, they were linked back to the interaction tasks and the relevant learning theories. For example, when students used Grammarly, their experiences mostly fit into the theme of guided or step-by-step support, which connects well with the principles of cognitivism. On the other hand, when students had open-ended, reflective conversations with ChatGPT, the insights leaned more towards themes like personal development and independence in learning—ideas that are closely associated with humanistic learning theory.

By the end of the analysis phase, each participant's experience had been classified along two dimensions: (1) interaction paradigm (e.g., directive, assistive, dialogic, co-constructive) and (2) associated learning theory (e.g., behaviorism, cognitivism, etc.). This two-axis mapping helped structure the findings and served as the basis for the synthesis presented in the next section.

## 8. Findings

From the interviews and observational notes, several recurring patterns emerged about how students interacted with various AI tools and made sense of their learning journey. The analysis led to five key themes: Feedback and Reinforcement, Cognitive Scaffolding, Dialogic Engagement, Personalization and Empathy, and Learning Agency. Each of these themes represents a major way in which students connected with the AI systems, and each one links closely with one or more well-known educational learning theories.

*8.1 Feedback and Reinforcement: Mapping to Behaviorism*

Participants who worked on math-related tasks through Khan Academy or Socratic AI often spoke about getting instant feedback and being able to correct their mistakes quickly. These tools offered clear answers or step-by-step hints, which helped students understand where they went wrong and try again. This kind of back-and-forth supported a trial-and-error approach. As one student shared:

"It felt like a drill—try, fail, get a hint, and try again. It reminded me of how I learned multiplication tables in school."

This type of interaction reflected core behaviorist principles, where learning happens through stimulus, response, and reinforcement. The AI tools acted like digital tutors—repeating instructions, correcting errors, and praising right answers—which helped students absorb facts and follow set procedures. The repetition and instant feedback made the learning process more structured and habit-forming.

*8.2 Cognitive Scaffolding: Mapping to Cognitivism*

Students who used Grammarly or ChatGPT for essay revision shared that these tools didn't just fix mistakes—they helped them think better and write more clearly. The AI pointed out grammar slips, suggested better sentence structures, and explained why a change was needed. One student shared:

"Grammarly didn't just fix the sentence—it showed me why it was wrong. I could see the logic."

These experiences align with cognitivist learning, where students engage in active information processing and develop internal understanding through structured guidance. The AI tools acted as mental support systems, aiding in focus, retention, and better grasp of concepts.

*8.3 Dialogic Engagement: Mapping to Constructivism*

In open-ended conversations with ChatGPT, students explored personal or academic issues such as stress management, ethical decision-making, or exam preparation. The AI responded with probing questions, summaries of perspectives, or advice grounded in general knowledge. Participants described this experience as "co-thinking" or "bouncing off ideas":

"ChatGPT didn't give me a yes or no answer. It felt like we were working it out together."

This kind of back-and-forth discussion fits well with constructivist learning theories, where understanding grows through reflection and meaningful exchange between the student and a supportive guide like the AI.

*8.4 Personalization and Empathy: Mapping to Humanism*

Many participants shared that interacting with the AI felt emotionally supportive or uplifting, particularly during the open-ended activity. Even if the responses weren't very specific, just feeling "heard" or having their thoughts acknowledged made them feel motivated and encouraged.

"It didn't feel like a robot. I felt like my point of view was heard. That made me want to continue."

These feelings resonate with the principles of humanistic learning theories, which focus on personal growth, emotional connection, and learner autonomy. In this context, the AI contributed to students' emotional well-being and helped nurture their inner motivation to learn.

*8.5 Learning Agency and Self-Perception: Cross-Theory Insight*

Across the different tasks, students' sense of control or agency shifted noticeably. In more structured or directive activities like solving math problems, some participants felt like they were simply following instructions, reflected in remarks such as:

"I was just clicking through until it said I was right."

In contrast, in dialogic and reflective tasks, students described higher ownership of learning:

"I chose what to ask and followed up how I wanted. That made it more mine."

These differences underscore how the nature of interaction influences students' feelings of control, interest, and how they see themselves as learners. While the behaviorist-style tasks were seen as quick and efficient, the constructivist and humanistic interactions led to more thoughtful engagement and emotional connection with the learning process.

Table 3. Summary of Findings: Mapping HAI Interactions to Learning Theories and Themes Table

| Task / Tool | Interaction Paradigm | Dominant Learning Theory | Thematic Insight | Student Experience Keywords |
|---|---|---|---|---|
| **Math problem-solving with Khan Academy / Socratic AI** | Directive, Feedback-Based | Behaviorism | Feedback & Reinforcement | Drill, repetition, correction, reinforcement |
| **Essay revision using Grammarly / ChatGPT** | Assistive, Scaffolded | Cognitivism | Cognitive Scaffolding | Structure, logic, explanation, mental models |
| **Open-ended dialogue with ChatGPT** | Dialogic, Co-Constructive | Constructivism | Dialogic Engagement | Reflection, idea building, co-thinking |
| **Personal exploration with ChatGPT** | Supportive, Empathetic | Humanism | Personalization and Empathy | Validation, emotional support, self-expression |
| **Across all tasks** | Varies by design | Cross-Theoretical | Learning Agency and Self-Perception | Ownership, control, dependency, engagement level |

## 9. Conclusion

This study took a closer look at how college students engage with AI tools in their learning journey, especially focusing on how these interactions connect with well-known educational theories. Through a mix of hands-on tasks and follow-up interviews, we tried to understand how tools like ChatGPT, Grammarly, and Khan Academy are being used in real learning situations. The aim was to explore whether these student-AI interactions reflect different teaching and learning styles—such as behaviorism, where feedback and repetition matter; cognitivism, which focuses on mental processes; constructivism, which emphasizes building knowledge through experience; and humanism, which values personal growth and emotional well-being. By linking AI usage to these frameworks, the study brings out a more grounded view of how intelligent systems are reshaping the learning experience for students in higher education.

The results of this study indicate that no single educational theory can capture the full range of student experiences while interacting with AI tools. Rather, different types of tasks and AI functionalities appear to activate varied learning processes. Tools that provided direct answers and reinforced correct attempts were more in tune with behaviorist principles. On the other hand, platforms that offered guided feedback and supported mental organization aligned well with cognitivist thinking. Interactions that were more conversational and tailored to individual needs reflected the ideals of constructivist and humanistic learning, encouraging deeper reflection and emotional connection. A recurring insight across all cases was the shifting sense of student agency—how much control or ownership learners felt they had over their own learning. This greatly influenced how they interpreted their role in an AI-supported educational setting.

This study attempts to fill an important gap in current educational research by presenting a theory-based and evidence-backed framework that connects AI-driven student interactions with established learning theories. While a lot of existing work tends to highlight the technical aspects of AI or its general influence on education systems, very few have taken a structured, theory-oriented approach to examine how Human-AI Interaction (HAI) actually plays out in learning environments. By linking students' actual behaviors and reflections to foundational educational theories, this research offers a useful perspective for teachers, curriculum designers, and EdTech professionals. It helps them think more deeply about how to create or refine AI tools in a way that truly supports meaningful and effective learning experiences.

This study brings to light the need for thoughtful design when using AI tools in the education sector. With more colleges and universities turning to smart tutoring systems, chatbots, and AI-driven content platforms, it is becoming essential to ensure that the design of these tools matches educational goals. It is not just about how well the AI works technically, but also about how students perceive and engage with it. The learning experience depends heavily on this interaction. Developers and educators, therefore, must work together to make sure AI tools are not only efficient but also ethical, meaningful, and truly helpful for learners. While this research offers valuable insights, it also has some limitations—like a small sample size, use of only a few AI tools, and being limited to one university. These open the door for future studies that can look into different academic levels, varied cultural contexts, or how long-term use of AI might affect students' learning habits and academic confidence. Overall, this study stresses the point that students are not just users of AI—they are learners whose journey must be supported by systems rooted in solid teaching principles. It urges a shift in focus from simply developing AI tools to designing them with the learner at the center, backed by theory and tested through real-world use.

## 10. Discussion

The findings of this study align with recent literature highlighting the diverse ways in which students engage with AI-based educational tools. The behaviorist patterns observed through reinforcement-based tasks echo the findings of Lo (2023) and Ali et al. (2023), who noted that immediate feedback enhances procedural learning. Similarly, the cognitivist themes identified in this study resonate with Mayer (2024) and Lombardi et al. (2024), emphasizing the role of AI tools in supporting cognitive organization and scaffolding. The constructivist and humanistic elements, evident in open-ended interactions, are consistent with Kasneci et al. (2023) and Tlili et al. (2023), who argued that dialogic AI environments foster learner autonomy and reflective thinking. Finally, the emphasis on learner agency observed here supports the connectivist perspective advanced by Noh and Lee (2020), underscoring the importance of distributed learning networks. Collectively, these findings reinforce the idea that AI tools can be pedagogically aligned with multiple learning theories, depending on their design and use context.

## 11. Limitations and Future Work

Although this study provides valuable perspectives on how students interact with AI tools within the framework of established learning theories, there are certain limitations that must be acknowledged. The participant group was drawn solely from undergraduate students at a single institution, which may limit how well the findings apply to learners from different educational backgrounds, age groups, or cultural settings. Additionally, the study focused on a limited number of AI tools and tasks. As a result, not all possible modes of student-AI interaction could be captured. The primary data was collected through self-reported interviews, which, while insightful, might be affected by participants' memory gaps or a tendency to present socially acceptable responses.

Looking ahead, future research should aim to include a broader and more diverse set of learners, including school students, postgraduate scholars, or professionals engaged in lifelong learning. It would also be helpful to explore a wider range of AI tools—such as adaptive platforms, virtual avatars, and immersive learning technologies—to capture different styles of interaction and pedagogical potential. Long-term studies could provide insights into how continuous use of AI tools influences a student's learning habits, academic confidence, and motivation. Moreover, the inclusion of multimodal data sources—like screen recordings, eye-tracking, or emotion sensors—could help researchers understand the deeper cognitive and emotional processes at play during AI-based learning sessions. Such approaches would not only strengthen the theoretical model developed in this study but also make it more responsive to real-world classroom needs.

**Acknowledgments**

The authors sincerely thank the students who participated in this study for generously sharing their time, experiences, and insights. Their reflections were invaluable in shaping the understanding presented in this research. The authors also acknowledge the support of the all institutions for providing the facilities required to conduct this study.

**Authors contributions**

Not applicable.

**Funding**

Not applicable.

**Competing interests**

Not applicable.

**Informed consent**

Obtained.

**Ethics approval**

The Publication Ethics Committee of the Macrothink Institute.

The journal's policies adhere to the Core Practices established by the Committee on Publication Ethics (COPE).

**Provenance and peer review**

Not commissioned; externally double-blind peer reviewed.

**Data availability statement**

The data that support the findings of this study are available on request from the corresponding author. The data are not publicly available due to privacy or ethical restrictions.

**Data sharing statement**

No additional data are available.

**Open access**

This is an open-access article distributed under the terms and conditions of the Creative Commons Attribution license (http://creativecommons.org/licenses/by/4.0/).

**Copyrights**

Copyright for this article is retained by the author(s), with first publication rights granted to the journal.

# Appendix A

## Structured Plan for Experiment

| Component | Details |
| --- | --- |
| Objective | To examine how undergraduate students interact with software agents (e.g., ChatGPT) across various learning tasks and how these interactions align with educational learning theories (Behaviorism, Cognitivism, Constructivism, Humanism, Connectivism). |
| Participants | 15–20 undergraduate students from diverse disciplines with basic digital literacy. |
| Setting | University computer lab or remote setup with internet-enabled devices. |
| Tools Used | ChatGPT or similar LLM-based AI, Google Docs for task submission, screen-recording tools for capturing interaction logs. |
| Duration | 45–60 minutes per participant session. |
| Pre-Task | Brief orientation about the experiment, consent form signing, and demographic information collection. |
| Tasks Given | 1. MCQ-based feedback task (Behaviorism) <br> 2. Step-by-step concept explanation (Cognitivism) <br> 3. Project brainstorming session (Constructivism) <br> 4. Reflection writing task (Humanism) <br> 5. Collaborative problem-solving using AI prompts (Connectivism) |
| Interaction Mode | Text-based conversation between student and AI agent; screen activity recorded. |
| Post-Task | Semi-structured interview for qualitative feedback. |
| Data Collection | Data Collection    Interaction transcripts, user reflections, and interview responses. |
| Data Analysis | Data Analysis    Thematic analysis to map interaction types with learning theories. |

**Appendix B**

**Semi-Structured Interview Guide**

**Purpose:** To explore students' subjective experience, perceived usefulness, engagement, and preferences regarding software-agent-based learning interactions.

**Section 1: Background**

1. Can you briefly describe your previous experience using AI tools (like ChatGPT)?
2. Have you used any AI-based learning platforms in your academic work before?

**Section 2: Interaction Experience**

1. How would you describe your interaction with the AI agent during the tasks?
2. Did you find the AI's responses helpful or supportive? Why or why not?
3. Were there moments you felt the AI understood your needs or adapted to you?

**Section 3: Task-Specific Feedback**

1. For the quiz/MCQ task (Behaviorism):
   - Did the feedback from the AI influence your understanding or motivation?
2. For the step-by-step explanation (Cognitivism):
   - How well did the AI help you build on your prior knowledge?
3. For the project brainstorming session (Constructivism):
   - Did interacting with the AI help you think creatively or develop your own ideas?
4. For the reflection task (Humanism):
   - Did the AI encourage personal expression or self-awareness?
5. For the collaboration task (Connectivism):
   - Did you feel like you were learning collaboratively with the AI?

**Section 4: Overall Reflection**

1. Which task felt most natural or engaging to you? Why?
2. Which task felt least effective or engaging? Why?
3. What learning theories or teaching methods do you feel this kind of AI supports best?
4. Would you be comfortable using this AI regularly in your coursework?
5. What improvements would you suggest for better educational use of such tools?